\documentclass[twocolumn,showpacs,preprintnumbers,amsmath,amssymb]{revtex4}
\usepackage[english]{babel}
\usepackage{graphicx,epsfig,psfig}
\usepackage{dcolumn}
\usepackage{bm}
\begin{document}
%
%
\def\bsig{\mbox{\boldmath $\sigma$}}                          
\def\bsig{\mbox{\boldmath $\Sigma$}}
\def\bgam{\mbox{\boldmath $\gamma$}}
\def\bgam{\mbox{\boldmath $\Gamma$}}
\def\bphi{\mbox{\boldmath $\phi$}}
\def\bphi{\mbox{\boldmath $\Phi$}}
\def\btau{\mbox{\boldmath $\tau$}}
\def\btau{\mbox{\boldmath $\Tau$}}
\def\btau{\mbox{\boldmath $\partial$}}
\def\Delc{{\Delta}_{\circ}}
\def\bp{\mid {\bf p} \mid}
\def\al{\alpha}
\def\bet{\beta}
\def\gam{\gamma}
\def\del{\delta}
\def\Del{\Delta}
\def\te{\theta}
\def\nua{{\nu}_{\alpha}}
\def\nui{{\nu}_i}
\def\nuj{{\nu}_j}
\def\nue{{\nu}_e}
\def\num{{\nu}_{\mu}}
\def\nut{{\nu}_{\tau}}
\def\2te{2{\theta}}
\def\chic#1{{\scriptscriptstyle #1}}
\def\chicl{{\chic L}}
\def\lam{\lambda}
\def\SU{SU(2)_{\chic W} \otimes U(1)_{\chic Y}}
\def\Lam{\Lambda}
\def\sig{\sigma}
\def\O{\Omega}
\def\o{\omega}
\def\s{\sigma}
\def\D{\Delta}
\def\d{\delta}
\def\df{\rm d}
\def\8{\infty}
\def\ld{\lambda}
\def\eps{\epsilon}
\def\ii{\'{\i}}  
\def\vsk{\vspace{0.52cm}} 
\def\vseq{\vspace{0.05cm}}  
\def\bea{\begin{eqnarray}} 
\def\eea{\end{eqnarray}} 
\def\ni{\noindent} 
\def\lb{\label} 
\def\nn{\nonumber}
\def\H{\cal H\it(t)}
\def\Hdiag{\cal H\it_D (t)}
\def\Oxy{\cal O_{\rm\chic{12}}\it}
\def\Oxz{\cal O_{\rm\chic{13}}\it}
\def\Oyz{\cal O_{\rm\chic{23}}\it}
\def\G{\rm\Gamma\it}
\def\Hhat{\widehat {\cal H}\it(t)}
\def\Htilde{\widetilde {\cal H}\it(t)}
\def\Om{\cal O\it_m (t)}
\def\Omhat{\widehat{\cal O \it}_m}
\def\Hhatb{\widehat {\cal H}_{\widehat B = 0}}
\def\Htildeb{\widetilde {\cal H}_{\widetilde B = 0}}
%
\newcommand{\be}{\begin{equation}}
\newcommand{\ee}{\end{equation}}
\newcommand{\ba}{\begin{array}}
\newcommand{\ea}{\end{array}}
\newcommand{\dis}{\displaystyle}
\newcommand{\alfad}{\frac{\dis \bar \alpha_s}{\dis \pi}}
\title{Magnus Expansion and Three-Neutrino Oscillations in Matter
\footnote{Presented at {\it Mexican School of Astrophysics} 
({\bf EMA}), Guanajuato, Mexico, July 31 - August 7, 2002.
Final version to appear in the {\bf Proceedings of IX Mexican Workshop 
on Particles and Fields} {\it ``Physics Beyond the Standard Model''}, 
Colima Col. M\'exico, November 17-22, 2003.} }
\author{Luis G. Cabral-Rosetti}
 \email{luis@nucleares.unam.mx ; lgcabral@ciidet.edu.mx}
\affiliation{Departamento de Posgrado,\\
Centro Interdisciplinario de Investigaci\'on y Docencia en 
Educaci\'on T\'ecnica (CIIDET),\\
Av. Universidad 282 Pte., Col. Centro, A. Postal 752, C. P. 76000,\\
Santiago de Queretaro, Qro., M\'exico.}%
\author{Alexis A. Aguilar-Arevalo}
\author{J. C. D'Olivo}
 \email{dolivo@nuclecu.unam.mx}
\affiliation{Instituto de Ciencias Nucleares, \\
             Departameto de F{\'\i}sica de Altas Energ{\'\i}as,\\
             Universidad Nacional Aut\'onoma de M\'exico (ICN-UNAM).\\
             Apartado Postal 70-543, 04510 M\'exico, D. F., M\'exico.}
\begin{abstract}
We present a semi-analytical derivation of the survival probability of
solar neutrinos in the three generation scheme, based on the Magnus 
approximation of the evolution operator of a three level system, and 
assuming a mass hierarchy among neutrino mass eigenstates. We have used 
an exponential profile for the solar electron density in our approximation. 
The different interesting density regions that appear throughout the 
propagation are analyzed. Finally, some comments on the allowed regions 
in the solar neutrino parameter space are addressed.
\end{abstract}
\maketitle
\section{Introduction}
The need to introduce the three generations of neutrinos into a computation
of the transition probabilities of these particles while traversing a medium 
has been recognised long ago \cite{need} to try to accomodate the observations
of different experiments studying neutrino oscillations. Analytical treatments
of three neutrino oscillations in matter with varying density in the three 
generation scheme have been studied in the past \cite{tres} aiming to deduce 
expressions for the oscillation probabilities of one type of neutrino into 
another. It has been shown, first in the case of two generations 
\cite{petcov_1,toshev}, and later in the case of three generations 
\cite{osland} that the differential equation describing the evolution of the 
neutrino state in an exponentialy varying density profile could be solved 
analytically in terms of confluent hypergeometric functions. Corrections to 
the mixing parameters in matter calculated as series expansions have been 
performed by Freund \cite{freund}, and a different perturbative analysis has 
been done by Narayan in \cite{narayan}. Global analyses of the recent 
experimental dathave been extensively studied both, in the two and three 
generations cases  \cite{concha}. In a previous paper by D'Olivo and Oteo 
\cite{dolivo_oteo} an approximate expression to the evolution operator using 
the Magnus expansion was found, only for the non-adiabatic regime in the 
exponentially varying density profile. In this paper we present a complete 
semi-analytical computation of the evolution operator for neutrinos in the 
same density profile, using the Magnus expansion approximation which will 
work propperly in the case of adiabatic and non-adiabatic evolution of the 
neutrino state. This paper describes some important results of the M. C. 
Thesis work presented by Luis G. Cabral-Rosetti in \cite{tesis}. Let $H$ 
denote the Hamiltonian of a quantum systen and $U=U(t,t_0)$ the time evolution
operator satisfying the Schr\"{o}dinger equation
\be
\lb{eq_sch}
i\hbar \;\frac{\partial}{\partial t} U =
HU \; , \;\;\; U(t_0,t_0) = I \; .
\ee
When $H$ is independent of time, or more generally when 
$[\int_{t_0}^{t}dt'H(t'),H(t)] = 0$, the solution of Eq.(\ref{eq_sch}) is 
formally $U={\rm Exp}[-i/\hbar\int_{t_0}^{t}dt'H(t')]$. Then it is natural 
to ask wether a solution of the form $U={\rm Exp}\Omega$ would always be 
possible. A method for finding such a {\it true} exponential solution 
(without time ordering) is supplied by the Magnus Expansion (ME) 
\cite{magnus}. The magnus operator $\Omega={\rm ln}U$ satisfies a 
differential equation which in turn is solved through a series expansion: 
$\Omega = \sum_{n=1}^{\infty}\Omega_n$, where each term $\Omega_n$ is of 
order $\hbar^{-n}$. The first two contributions are explicitly given by
$$
\Omega_1 = -\frac{i}{\hbar}\int_{t_0}^{t}dt_1H(t_1)\; ,  
$$
\be
\Omega_2 =  -\frac{i}{2\hbar^2}\int_{t_0}^{t}dt_1\int_{t_0}^{t_1}dt_2
[H(t_1),H(t_2)]\; .
\ee
Recursive methods to obtain the succesive terms have been extensively worked 
out in the literature \cite{burum}. Because of the anti-Hermitian character 
of every $\Omega_n$, each approximate time-evolution operator obtained as 
$U\approx U_k = {\rm Exp}(\sum_{n=1}^{k}\Omega_n)$ will be unitary. Here we 
use the first Magnus approximant to obtain (approximate) analytical solutions 
to the problem of 3 neutrinos oscillations in a medium with varying density as 
the Sun.
\begin{figure}[htp]
\centering  
\rotatebox{-90}{\scalebox{0.30}{\includegraphics{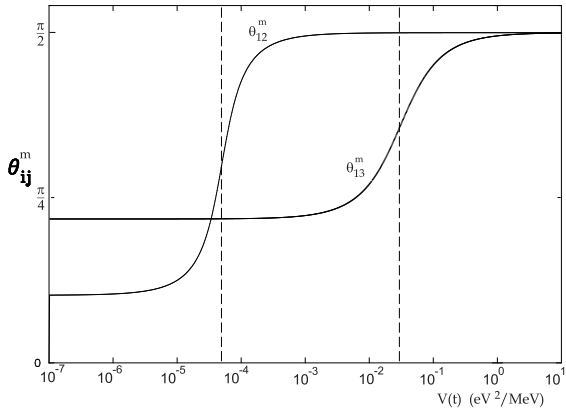}}} 
\caption{\small Behaviour of the matter mixing angles as functions
of the effective potential $V(t)$.\normalsize}
\label{angulos} 
\end{figure} 
\section{Survival Probability for electron Neutrinos}
From the dependency of ${\rm sin}\; 2{\te}_{\chic {12}}^{\chic {m}}(t)$, 
and ${\rm sin}\; 2{\te}_{\chic {13}}^{\chic {m}}(t)$ on $V(t)$ (see Ref. 
\cite{tesis} and \cite{nosotros} for all details), we can distinguish five 
interesting regions where the oscillations behave differently.
\begin{description}
\item{(1)} {\bf Region of Extreemely Low Density}\\ \\
If $0 \leq V(t) \ll {\Del}_{\chic {21}}$, both mixing angles,
${\te}_{\chic {12}}^{\chic m}(t)$, and ${\te}_{\chic {13}}^{\chic m}(t)$,
are close to their values in vacuum (see Fig.(\ref{angulos})), this is 
${\te}_{\chic{12}}^{\chic m}(t) \rightarrow {\te}_{\chic {12}}$ ; 
${\te}_{\chic{13}}^{\chic m}(t) \rightarrow {\te}_{\chic {13}}$, 
giving
\be
{\cal P}_{\al}(t) \rightarrow c_{\chic {12}}^{\chic 2}\; c_{\chic
{13}}^{\chic 2} \;, \;\;{\rm and} \;\;\;
{\cal P}_{\gam}(t) \rightarrow s_{\chic {13}}^{\chic 2}\ ,
\label{eq92}
\ee
\ni
leading us to the well known result of vacuum oscillations of three 
neutrinos
\be
\big<{\cal P}({\nue} \rightarrow {\nue})\big> = c_{\chic {12}}^{\chic
4}\; c_{\chic {13}}^{\chic 4} + s_{\chic {12}}^{\chic 4}\; c_{\chic
{13}}^{\chic 4} + s_{\chic {13}}^{\chic 4}\ ,
\label{eq93}
\ee
\item{(2)} {\bf Low Density Resonance Region}\\ \\
For $V(t) \approx {\Del}_{\chic {21}}$, ${\te}_{\chic {13}}^{\chic
m}(t)$ is still close to its vacuum value
(${\te}_{\chic {13}}^{\chic m}(t) \approx {\te}_{\chic {13}}$),
while ${\te}_{\chic {12}}^{\chic m}(t)$ is at the 
{\it low density resonance}, i.e. ${\te}_{\chic {12}}^{\chic m}(t)
\ \approx {{{\pi}\over {4}}}$, making
\bea
{\cal P}_{\al}(t) &\rightarrow& {\rm cos}^{2}\; {\Pi}(t)\ {\rm cos}^{2}\;
{\te}_{\chic {12}}^{\chic m}(t_{\chic 0})\; {\rm cos}^{2}\; {\te}_{\chic
{13}}^{\chic m}(t_{\chic 0}) \nn \\
&&+ {\rm sin}^{2}\; {\Pi}(t)\ {\rm sin}^{2}\;
{\te}_{\chic {12}}^{\chic m}(t_{\chic 0})\; {\rm cos}^{2}\; {\te}_{\chic
{13}}^{\chic m}(t_{\chic 0}) \; ,
\label{eq94}
\eea
and
\be
{\cal P}_{\gam}(t) \rightarrow {\rm sin}^{2}\; {\te}_{\chic {13}}^{\chic
m}(t_{\chic 0})\ , 
\label{eq95}
\ee
\ni which leads to the expression
\be
\big<{\cal P}_{({\nue} \rightarrow {\nue})}\big> = {{1}\over {2}} +
{{1}\over {2}}\Big(1 - 2{\cal P}_{c}^{\; l}\Big)\;
{\rm cos}\;2{\te}_{\chic {12}}^{m}(t_{\chic 0})\; {\rm cos}\; 
2{\te}_{\chic {12}}\ ,
\label{eq96}
\ee
\ni 
where
\be
{\rm cos}\; 2{\te}_{\chic {12}}^{m}(t_{\chic 0}) = {{V_{\chic l} 
- V(t_{\chic 0})}\over {\sqrt{(V_\chic{l} - V(t_{\chic 0}))^2 + B_{\chic
l}^2}}}\ ,
\label{eq97}
\ee
\ni
and
\be
{\cal P}_{c}^{\; l} = {\rm sin}^{\chic 2}\Big[\Big({\te}_{\chic
{12}}^{\chic m}(t_{\chic 0}) - {\te}_{\chic {12}}^{\chic m}(T)\Big)
{\hbox{Exp}{\ }}{(-\ {\kappa}_{\chic l})}\Big]\ .
\label{eq98}
\ee
\begin{figure}[htp]
\rotatebox{-90}{\scalebox{0.30}{\includegraphics{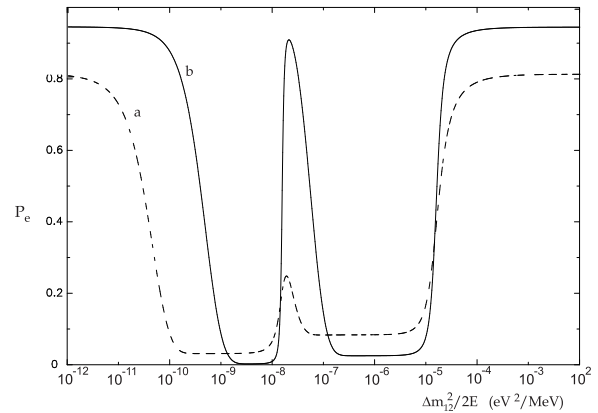}}} 
\caption{\label{prob2} The survival probability for an electron neutrino
comming from the Sun vs. ${\del}\,
m^{2}_{\chic{21}}/2E$, with the parameters 
(a) ${\rm sin}^{2}\; 2{\te}_{\chic {12}} = 0.1$,
${\rm sin}^{2}\; 2{\te}_{\chic {13}} = 0.01$ y $R = 10^{3}$; 
(b) ${\rm sin}^{2}\; 2{\te}_{\chic {12}} = 0.3$,
${\rm sin}^{2}\; 2{\te}_{\chic {13}} = 0.1$ y $R = 10^{3}$. }
\end{figure}

The quantity ${\cal P}_{c}^{\; l}$ represents the tramsition probability
between the states ${\nu}_{\chic {2m}}$ y ${\nu}_{\chic{1m}}$, 
and has been derived by D'Olivo \cite{D'Olivo3} in the case
of oscillations between two neutrino species. 
The adiabatic result can be recovered by seting ${\cal P}_{c}^{\; l} = 0$.  
For ${\kappa}_{\chic l} \gg 1$, ${\cal P}_{c}^{\; l}$ is exponentially 
supressed as expected in the asimptotic regime. On the other hand,
for ${\kappa}_{\chic l} < 1$ there are significant corrections to the
adiabatic approximation which reduce the effect of the resonant transition.
Eq.(\ref{eq96}) was first obtained by Parke \cite{Parke4} for two species
using the Landau-Zener approximation for the crossing probability: ${\cal
P}_{c}^{\; l} = Exp\ \big(- {{\pi}\over {2}}{\kappa_{\chic l} }\big)$. In 
the extreeme non adiabatic case, ${\Del}_{\chic {21}} \rightarrow 0$, making 
$\kappa_\chic{l}\rightarrow 0$, and from Eq. (\ref{eq97}) it follows that 
${\te}_{\chic {12}}^{\chic m}(t_{\chic 0}) \rightarrow {{\pi}\over {2}}$, 
making ${\cal P}_{c}^{\; l} = {\rm cos}^{2}\; {\te}_{\chic {12}}$. Introducing
this value of ${\cal P}_{c}^{\; l}$ in Eq.(\ref{eq96}) we recover the vacuum 
result for two neutrinos:
\be
\big<{\cal P}_{({\nue} \rightarrow {\nue})}\big> = 1 - {{1}\over {2}}\;
{\rm sin}^{2}\; 2{\te}_{\chic 12} \ ,
\label{eq99}
\ee
This should be contrasted with the result 
\be 
\big<{\cal P}({\nue} \rightarrow {\nue})\big> = {\rm cos}^{2}\; 
{\te}_{\chic {12}}\ , 
\label{eq100}
\ee 
predicted by the Landau-Zener formula (and, in general, by any
result derived from the Dykhne's formula), which deviates from the
correct limit given by Eq.(\ref{eq99}), when ${\te}_{\chic {12}}$ 
is large. The correct value for the extreeme nonadiabatic case 
has been derived before, under the assumption that the transition 
between the adiabatic eigenstates occurs instantaneously 
at the time $t = t_{\chic l}$ \cite{Kim2}. 

\begin{figure}[htp]
\rotatebox{-90}{\scalebox{0.30}{\includegraphics{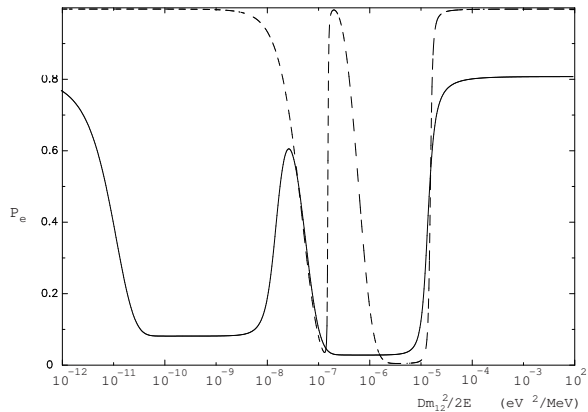}}} 
\caption{\label{prob3} The survival probability for an electron neutrino
comming from the Sun vs. ${\del}\,
m^{2}_{\chic{21}}/2E$, with the parameters 
(a) ${\rm sin}^{2}\; 2{\te}_{\chic {12}} = 0.1$,
${\rm sin}^{2}\; 2{\te}_{\chic {13}} = 0.001$ and $R = 100$.
(b) ${\rm sin}^{2}\; 2{\te}_{\chic {12}} = 0.1$,
${\rm sin}^{2}\; 2{\te}_{\chic {13}} = 0.3$ and $R = 10^{3}$.}
\end{figure}
However, in this case 
the corresponding result for ${\cal P}_{c}^{\; l}$ approaches  
${\rm cos}^{2}\; {\te}_{\chic {12}}$ as $({{\del}\; m^{\chic 2}\over E})^{2}$ 
instead of linearly, as in Eq.~(\ref{eq98}). Non adiabatic effects start 
to become important in this region when ${\kappa}_{\chic l}$ is comparable 
to  1, whenever the neutrinos cross the  {\it low density resonance}. 
If $E < {{{\del}\; m_{\chic {21}}^{\chic 2}}\over {2\; V(t_{\chic 0})}}\,
{\rm cos}\; 2{\te}_{\chic {12}}$, ${\cal P}_{c}^{\; l} = 0$, and the 
propagation of neutrinos will be adiabatic for ${\kappa}_{\chic l} < 1$. 
For this reason, the asimptotic exponential expression 
($\hbar \rightarrow 0$) for ${\cal P}_{c}^{\; l}$, must be modified by hand 
to consider this situation. An {\it effective} way of implementing such 
modification is to multiply ${\cal P}_{c}^{\; l}$ by a step function 
${\Theta}(V(t_{\chic 0}) - {\Del}_{\chic {21}}\, cos\; 2{\te}_{\chic{12}})$, 
in such a way that the transition probability vanishes if a neutrino is 
produced after the resonance. 
It is worthnoting that such modification is not necessary with the Magnus 
result Eq.~(\ref{eq98}) given that, as a function of 
${{\del}\; m_{\chic {21}}^{\chic 2}\over {2\; E}}$, the difference 
${\te}_{\chic {12}}^{\chic m}(t_{\chic 0}) - {\te}_{\chic {12}}^{\chic m}(T)$ 
behaves as a continuous step.\\

\item{(3)} {\bf Intermediate Density Region}\\ \\
For ${\Del}_{\chic {21}} \ll V(t) \ll {\Del}_{\chic {31}}$, the mixing angles
in matter are ${\te}_{\chic {12}}^{\chic
m}(t) \approx {{\pi}\over {2}}$, and  ${\te}_{\chic {13}}^{\chic m}(t)
\approx {\te}_{\chic {13}}$, making
\be
{\cal P}_{\al}(t) \approx c_{\chic {12}}^{\chic 2}\; c_{\chic
{13}}^{\chic 2} \; , 
\label{eq101}
\ee
\ni 
and
\be
{\cal P}_{\gam}(t) \approx s_{\chic {13}}^{\chic 2}\ ,
\label{eq102}
\ee
\ni giving the result shown in Eq.(\ref{eq93}), provided there exists a
clear separation between the resonance regions: 
$\Del_\chic{31}\gg\Del_\chic{21}$.\\

\begin{figure}[htp]
\rotatebox{-90}{\scalebox{0.30}{\includegraphics{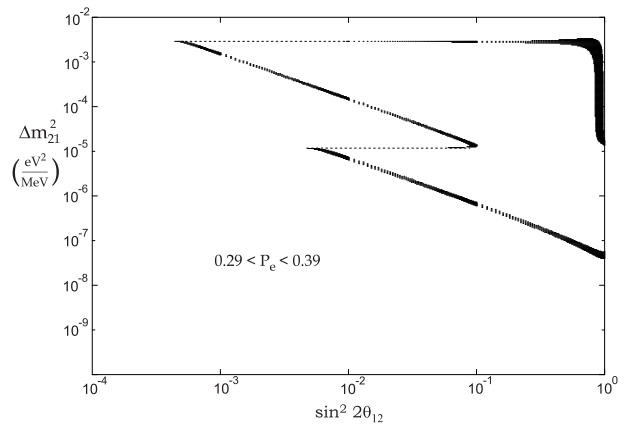}}} 
\caption{\label{sno_sk} The $1\sigma$ allowed region in the $\Delta_{21}$vs
${\rm sin}^2\;2\theta_{12}$ plane corresponding to the range
$0.25<P_e<0.43$ predicted by the combined results of SNO and Super Kamiokande.
Notice the double MSW triangle structure.}
\end{figure}
\item{(4)} {\bf High Density Resonance Region}\\ \\
When $V(t) \approx {\Del}_{\chic {31}}$, the mixing angle ${\te}_{\chic
{12}}^{\chic m}(t)$ is approximately equal to ${{\pi}\over {2}}$, while
${\te}_{\chic {13}}^{\chic m}(t)$ is at its {\it high density 
resonance} value of ${\te}_{\chic {13}}^{\chic m}(t)\ \approx
{{{\pi}\over {4}}}$. In this case
\be
{\cal P}_{\al}(t) \rightarrow {\rm cos}^{2}\; {\te}_{\chic {13}}^{\chic
m}(t_{\chic 0})\; {\rm cos}^{2}\; {\te}_{\chic {12}}^{\chic m}(t_{\chic 0}) 
\; ,
\label{eq103}
\ee
\ni
and
\be 
{\cal P}_{\gam}(t) \rightarrow cos^{2}\; {\Pi}(t)\; sin^{2}\;
{\te}_{\chic {13}}^{\chic m}(t_{\chic 0})\ ,
\label{eq104}
\ee
leading to 
\be
\big<{\cal P}_{({\nue} \rightarrow {\nue})}\big> = {{1}\over {2}} +
{{1}\over {2}}\Big(1 - 2{\cal P}_{c}^{\; h}\Big)\; {\rm cos}\; 2{\te}_{\chic
{13}}^{\chic m}(t_{\chic 0})\; {\rm cos}\; 2{\te}_{\chic {13}}
\label{eq105}
\ee
\ni
where
\be
{\rm cos}\; 2{\te}_{\chic {13}}^{\chic m}(t_{\chic 0}) = {{V_{\chic h} -
V(t_{\chic 0})}\over {\sqrt{(V(t_{\chic 0}) - V_{\chic h})^{\chic 2} +
B_{\chic h}^{\chic 2}}}}\ ,
\label{eq106}
\ee
\\
and
\be
{\cal P}_{c}^{\; h} = 
{\rm sin}^{\chic 2}\Big[\Big({\te}_{\chic {13}}^{\chic m}(t_{\chic 0}) 
- {\te}_{\chic {13}}^{\chic m}(T)\Big){\hbox{Exp}{\ }}
{(-\ {\kappa}_{\chic h})}
\Big]\ ,
\label{eq107}
\ee
\ni
which  again implies the two neutrino oscillations result.
The crossing probability  between the instantaneous eigenstates
${\nu}_{\chic {3m}}$  and ${\nu}_{\chic {2m}}$ given by
${\cal P}_{c}^{\; h}$ is perfectly analogue to that studied in the
low density resonance region. We can recover the adiabatic case if we set 
${\cal P}_{c}^{\; h} = 0$, and  for ${\kappa}_{\chic h} \gg 1$, 
${\cal P}_{c}^{\; h}$ is exponentially supressed. The extreeme nonadiabatic 
case requires  ${\Del}_{\chic {31}} \rightarrow 0$, and from Eq.~(\ref{eq97}) 
we have ${\te}_{\chic {13}}^{\chic m}(t_{\chic 0}) \rightarrow {{\pi}\over 
{2}}$ leading to  ${\cal P}_{c}^{\; h} = {\rm cos}^{2} \; {\te}_{\chic {13}}$. 
All these observations will lead us to an expression of the form of 
Eq.(\ref{eq99}) with ${\te}_{\chic {12}}$ replaced by ${\te}_{\chic {13}}$. 
Non adiabatic effects become important when ${\kappa}_{\chic h}\approx 1$
provided the neutrino crosses the {\it High density resonance}. If $E <
{{{\del}\; m_{\chic {31}}^{\chic 2}}\over {2\; V(t_{\chic 0})}}\,
({\rm cos}\; 2{\te}_{\chic {13}} - R^{-\; 1}\; {\rm sin}^{\chic 2}\, 
{\te}_{\chic 12}\; {\rm cos}\, 2{\te}_{\chic {13}})$, with 
$R=\Del_\chic{31}/\Del_\chic{21}$, no transitions between the instantaneous 
eigenstates can occur (${\cal P}_{c}^{\; h} = 0$), and the propagation 
is adiabatic for ${\kappa}_{\chic h} < 1$. Again, the difference 
${\te}_{\chic {13}}^{\chic m}(t_{\chic 0}) - {\te}_{\chic {13}}^{\chic m}(T)$ 
behaves as a continuous step giving ${\cal P}_c^{\;h}$ the appropiate 
behaviour.\\
\begin{figure}[htp]
\rotatebox{-90}{\scalebox{0.30}{\includegraphics{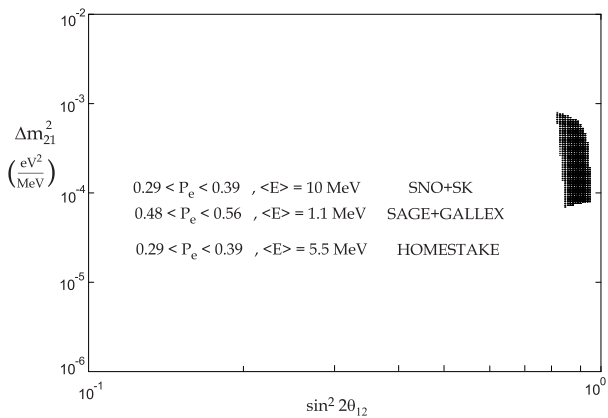}}} 
\caption{\label{allowed} Overlap of the allowed regions for the experiments
SAGE$+$GALLEX ($0.46<P<0.66$) and SNO$+$Super-Kamiokande ($0.25<P<0.43$). 
$R=S=250$.}
\end{figure}

\item{(5)} {\bf Extreemely High Density Region}\\ \\
For $V(t) \gg {\Del}_{\chic {31}}$ the $(\nue)$ oscillations 
are strongly supressed due to the fact that
${\te}_{\chic {12}}^{\chic m}(t) \approx {{\pi}\over {2}}$, and
${\te}_{\chic {13}}^{\chic m}(t) \approx {{\pi}\over {2}}$, making
\be
{\cal P}_{\al}(t) \rightarrow c_{\chic {12}}^{\chic 2}\; c_{\chic
{13}}^{\chic 2}
\label{eq171}
\ee
\ni
and
\be
{\cal P}_{\gam}(t) \rightarrow s_{\chic {13}}^{\chic 2}\ ,
\label{eq172}
\ee
\ni giving the result of Eq.(\ref{eq93})

\be
\big<{\cal P}({\nue} \rightarrow {\nue})\big> = c_{\chic {12}}^{\chic
4}\; c_{\chic {13}}^{\chic 4} + s_{\chic {12}}^{\chic 4}\; c_{\chic
{13}}^{\chic 4} + s_{\chic {13}}^{\chic 4}\ ,
\label{eq173}
\ee

\end{description}

Plots of the $\nue$ survival probability  $\big<P_{({\nue}
\rightarrow {\nue})}\big>$ as a function of ${{\del}\,{m}^{2}\over {2E}}$
are shown in Figs.2--3, for different  values of 
${\rm sin}^{2}2{\te}_{\chic {12}}$, ${\rm sin}^{2}2{\te}_{\chic {13}}$, 
and $R$. We use the exponential profile \cite{bahcall10} 
$N_{e}(r) = 245\ Exp(-\ 10.54\ r/ R_{\odot})\ N_{\chic {Avo}}\ cm^{-\ 3}$, 
where $N_{\chic {Avo}}$ is the Avogadro's number, r is the radial distance 
measured from the center of the Sun, and $R_{\odot}$ is the solar radius  
($R_{\odot} = 6.96 \times 10^{5}\ Km$). Except for those regions close to 
the center or the surface, this is a good approximation of the electron 
density in the Sun. We further assume that the ${\nue}$ are produced at 
$r_{\chic 0} = 0.08635\ R_{\odot}$, where $N_{e}$ is the central electron 
density predicted by the Standard Solar Model (SSM). In Fig.\ref{sno_sk} 
we show the allowed region in the $\Delta_{21}$ vs ${\rm sin}^22\theta_{12}$ 
plane obtained by simply plotting the points in the plane wit survival 
probabilities lying within the $1\sigma$ range of values extracted from 
recent joint analyses of the results of {\bf  SNO} and {\bf Super 
Kamiokande} \cite{implications1}. To produce this region we used the 
estimated value $P_e=(0.34\pm 0.05)$ for the survival probability, and an 
average neutrino energy of 5 MeV. We computed similar regions using the 
survival probabilities estimated by the experiments {\bf SAGE}, 
{\bf GALEX} and {\bf HOMESTAKE}, taking the average energies for this 
experiments as in \cite{ohlsson}, and looked for an overlap of these regions 
in the parameter plane (see Fig.~5). No intention to give  statistical 
significance to this region exists, but only it is shown that our result 
it is consistent with those achieved by rigurous analyses.  

\begin{acknowledgments}
This work has been supported in part by {\bf DGAPA-UNAM} under Grant 
{\bf PAPIIT} proyect {\tt No. IN109001} and in part by {\bf CoNaCyT} 
(M\'exico) under Grant {\tt No. I30307-E}.
\end{acknowledgments}


\begin{thebibliography}{99}
\bibitem{need}{K. S. Babu, Jogesh C. Pati, Frank Wilczec, 
{\it Phys. Lett. B} {\bf 359}, 351 (1995); Chritian Y. Cardall, 
George M. Fuller, {\it Phys. Rev. D} {\bf 53}, 8 (1996); H. Schlattl, 
{\it Phys. Rev. D} {\bf 64}, 013009 (2001); G. L. Fogli and E. Lisi 
{\it Phys. Rev. D} {\bf 54}, 3667 (1996); S. Goswami, 
{\it Phys. Rev. D} {\bf 55}, 2931 (1997); B. Armbruster {\it et. al}, 
{\it Phys. Rev. C} {\bf 57} 3414 (1998); C. Giunti, C. W. Kim, and M. 
Monteno, {\it Nucl. Phys. B} {521} (1998).}
\bibitem{tres}{T. K. Kuo and J. Pantaleone, {\it Phys.Rev.Lett} {\bf 57},
1805 (1986); C. W. Kim and W. K. Sze, {\it Phys.Rev. D} {\bf 35}, 1404 
(1987); G. L. Fogli, E. Lisi, and D. Montanino, {\it Phys. Rev. D} 
{\bf 49}, 3626 (1994); E. Torrente Lujan, {\it ibid.} {\bf 53}, 4030 
(1996); T. Sakai, O. Inagaki, and T. Teshima, {\it Int. J. Mod. Phys. A} {\bf 12}. 1953 ((1999); T. Ohlsson and H. Snellman, {\it J.Math.Phys.} 
{\bf 41}, 2768 (2000); G. L. Fogli, E. Lisi, D. Montanino, and A. Palazzo, 
{\it Phys. Rev. D} {\bf 62}, 013002 (2000).}
\bibitem{petcov_1}{S.T. Petcov, {\it Phys. Lett. B} {\bf 214}, 2 (1988).}
\bibitem{toshev}{S. Toshev, {\it Phys. Lett. B} {\bf 196}, 02 (1987).}
\bibitem{osland}{Per Osland, T. T. Wu, {\it Phys. Rev. D} {\bf 62}, 
013008 (2000).}
\bibitem{freund}{M. Freund, {\it Phys. Rev. D} {\bf 64}, 053003  (2001).}
\bibitem{narayan}{M. Narayan, M. V. N. Murthy, G. Rajasekaran, and S. Uma 
Sankar, {\it Phys. Rev. D} {\bf 53}, 2819 (1996).}
\bibitem{concha}{J. Bahcall, M. C. Gonzalez-Garcia, C. Pe\~na-Garay, 
hep-ph/0204314, (2002); J. Bahcall, M. C. Gonzalez-Garcia, C. Pe\~na-Garay, 
JHEP {\bf 0204}:014 (2002); J. Bahcall, M.C. Gonzalez-Garcia, C. 
Pe\~na-Garay, JHEP {\bf 0108}:014 (2001).}
\bibitem{dolivo_oteo}{J. C. D'Olivo and J. A. Oteo, {\it Phys. Rev. D}
{\bf 54}, 1187 (1996).}
\bibitem{tesis}{Luis G. Cabral-Rosetti, M. C. Thesis: {\it ``Tratamiento 
Anal{\'\i}tico para las Oscilaciones de Tres Neutrinos en Materia''}, 
Facultad de Ciencias de la Universidad Nacional Autonoma de M\'exico 
({\bf FC-UNAM}), M\'exico D. F., October (1994).}
\bibitem{magnus}{W. Magnus, {\it Commun. Pure Appl. Math.}, {\bf 7},
649 (1954).}
\bibitem{burum}{D. P. Burum, {\it Phys. Rev. B} {\bf 24}, 3684 (1981); 
W. R. Salzman, {\it J. Chem. Phys} {\bf 82}, 822 (1985); S. Klarsfeld and 
J. A. Oteo, {\it Phys. Rev. A} {\bf 39}, 3270 (1989).}
\bibitem{nosotros}{Alexis A. Aguilar-Arevalo, L. G. Cabral-Rosetti and 
J. C. D'Olivo in preparation.}
\bibitem{D'Olivo3}{J. C. D'Olivo, {\it Phys. Rev.}, {\bf D45}, 924
(1992).}\\
{J. C. D'Olivo and J. A. Oteo, {\it Phys. Rev.}, {\bf D42}, 256
(1990).}
\bibitem{Parke4}{W. C. Haxton, {\it Phys. Rev. Lett.} {\bf 57}, 1271
(1986).}\\ 
{S. J. Parke, {\it ibid.} {\bf 57}, 1275 (1986).}
\bibitem{Kim2}{C. K. Kim, S. Nussinov, and W. K. Sze, {\it Phys.
Lett.} {\bf B 184}, 403 (1987); {\it Phys. Rev.} {\bf D35}, 4014
(1987).} 
\bibitem{bahcall10}{J. Bahcall, S. Basu, M. H. Pissoneault, 
{\it Astrophys. J.}, {\bf 55}, 990 (2001).}
\bibitem{implications1}{G. L. Fogli, E. Lisi, D. Montanino, and, A. Palazzo, 
{\it Phys. Rev. D} {\bf 64}, 093007 (2001); A. Bandyopadhyay, S. Choubey,
S. Goswami, and, D.P. Roy, {\tt hep-ph 0204286} (2002).} 
\bibitem{ohlsson}{T. Ohlsson ans H. Snellman, {\it Phys. Rev. D} 
{\bf 60}, 093007 (2001).} 

\end{thebibliography}
\end{document}